\documentclass[aps,pra,twocolumn,10pt,amsmath,superscriptaddress,longbibliography,amssymb]{revtex4-1}
\usepackage{mathrsfs}
\usepackage{graphicx}
\usepackage{amsmath}
\usepackage{CJK}
\usepackage{amssymb}
\usepackage{amsfonts}
\usepackage{color}
\usepackage{multirow}

\usepackage{dcolumn}
\usepackage{bm}
\usepackage[pdfstartview=FitH, CJKbookmarks=true, bookmarksnumbered=true, bookmarksopen=true, colorlinks=true, pdfborder=001, citecolor=blue, linkcolor=blue, linktocpage=true] {hyperref}
\newcommand{\nn}{\nonumber}

\begin{document}

\preprint{APS/123-QED}
\title{Bloch oscillations of multi-magnon excitations in a Heisenberg XXZ chain}

\author{Wenjie Liu}
\affiliation{Laboratory of Quantum Engineering and Quantum Metrology, School of Physics and Astronomy, Sun Yat-Sen University (Zhuhai Campus), Zhuhai 519082, China}
\affiliation{State Key Laboratory of Optoelectronic Materials and Technologies, Sun Yat-Sen University (Guangzhou Campus), Guangzhou 510275, China}

\author{Yongguan Ke}
\affiliation{Laboratory of Quantum Engineering and Quantum Metrology, School of Physics and Astronomy, Sun Yat-Sen University (Zhuhai Campus), Zhuhai 519082, China}
\affiliation{State Key Laboratory of Optoelectronic Materials and Technologies, Sun Yat-Sen University (Guangzhou Campus), Guangzhou 510275, China}

\author{Li Zhang}
\affiliation{Laboratory of Quantum Engineering and Quantum Metrology, School of Physics and Astronomy, Sun Yat-Sen University (Zhuhai Campus), Zhuhai 519082, China}
\affiliation{State Key Laboratory of Optoelectronic Materials and Technologies, Sun Yat-Sen University (Guangzhou Campus), Guangzhou 510275, China}

\author{Chaohong Lee}
\altaffiliation{lichaoh2@mail.sysu.edu.cn}

\affiliation{Laboratory of Quantum Engineering and Quantum Metrology, School of Physics and Astronomy, Sun Yat-Sen University (Zhuhai Campus), Zhuhai 519082, China}
\affiliation{State Key Laboratory of Optoelectronic Materials and Technologies, Sun Yat-Sen University (Guangzhou Campus), Guangzhou 510275, China}
\affiliation{Synergetic Innovation Center for Quantum Effects and Applications, Hunan Normal University, Changsha 410081, China}

\date{\today}

\begin{abstract}
  The studies of multi-magnon excitations will extend our understandings of quantum magnetism and strongly correlated matters.
  Here, by using the time-evolving block decimation algorithm, we investigate the Bloch oscillations of two-magnon excitations under a gradient magnetic field.
  Through analyzing the symmetry of our Hamiltonian, we derive a rigorous and general relation between ferromagnetic and anti-ferromagnetic systems.
  Under strong interactions, in addition to free-magnon Bloch oscillations, there appear fractional bounded-magnon Bloch oscillations which can be understood by an effective single-particle model.
  To extract the frequencies of Bloch oscillations and determine the gradient of magnetic field, we analyze the fidelity and the sub-standard deviation in both time and frequency domains.
  Our study not only explores the interaction-induced Bloch oscillations of multi-magnon excitations, but also provides an alternative approach to determine the gradient of magnetic field via ultracold atoms in optical lattices.
\end{abstract}

\maketitle

\section{Introduction}
Heisenberg spin chain, a paradigmatic model in many-body physics, is benefit to study collective excitations and low-energy properties of quantum magnets.
Particularly, the elementary aspects of quantum magnetism can be well described by spin excitations.
The spin-wave theory provides a fundamental insight that magnons are the quasi-particle excitations over the ferromagnetic ground states~\cite{MWortis1963,MTakahashi1971}.
The Bethe ansatz has predicted the existence of magnon bound states (BSs) in Heisenberg chains~\cite{HABethe1931}.
It renders an attractive research subjects to identify the signatures of magnons~\cite{JSCaux2005,RGPereira2008,MKohno2009,AImambekov2012}.
The quench dynamics is considered as an effective way to probe magnon BSs~\cite{MGanahl2012,WLiu2014}.
It has demonstrated that ultracold atomic ensembles offer an ideal platform to simulate spin excitations~\cite{MLewenstein2007,APolkovnikov2011}.
In particular, single-magnon excitations and multi-magnon BSs have been observed in cold atom experiments~\cite{TFukuhara20131,TFukuhara20132}.

On the other hand, if a constant force is applied, a quantum particle in a periodic potential will undergo Bloch oscillations (BOs)~\cite{FBloch1929,CZener1934}.
The BOs have been directly observed with ultracold atoms~\cite{MBenDahan1996,BPAnderson1998,ZAGeiger2018}.
In multi-particle systems, inter-particle interaction will have an huge influence on the BOs.
For strongly interacting few-body system, novel fractional BOs can arise at a double (or integer multiple) Bloch frequency that of single-particle BOs~\cite{RKhomeriki2010}.
Fractional BOs have been studied in various systems, such as photonic systems~\cite{GCorrielli2013,SLonghi2011}, cold atom systems ~\cite{KWinkler2006,SFolling2007,PMPreiss2015}, electronic systems~\cite{WSDias2007}, etc.
In addition, BOs of magnetic solitons are theoretically studied in the spin chain systems~\cite{HBBraun1994,JKyriakidis1998}.
Similar to the BOs of an electron in a static electric field, single-magnon dynamics in spin chains subjected to a gradient magnetic field is examined~\cite{VVGann2010,HZhang2014,YAKosevich2013}.
However, the BOs of multi-magnon excitations are still unclear.
In particular, how to characterize and extract the novel effects induced by the inter-magnon interactions?

This paper aims to explore two-magnon dynamics in the Heisenberg spin chain under a gradient magnetic field.
We provide a dynamical symmetry analysis to explore the relation between ferromagnetic and anti-ferromagnetic systems.
By analyzing the spin distributions and longitudinal spin-spin correlations, we track the dynamical difference for different interactions via using the time-evolving block decimation (TEBD) algorithm~\cite{GVidal2003,GVidal2004}.
From the time-evolution of spin distributions and longitudinal spin-spin correlations, we find the dynamical signature from free-magnon BOs to bounded-magnon BOs.
We also calculate the fidelity and sub-standard deviation to extract the multi-frequency BOs and determine the gradient of magnetic field.

This paper is organized as follows.
In Sec.~\ref{Sec2}, we describe the Heisenberg XXZ chain within a gradient magnetic field and analyze its symmetry.
In Sec.~\ref{Sec3}, we simulate the BOs of two-magnon excitations via the TEBD algorithm.
In Sec.~\ref{Sec4}, we calculate the fidelity and the sub-standard deviation.
In Sec.~\ref{Sec5}, we give a brief summary and discussions.

\section{magnon excitations and their dynamical symmetry} \label{Sec2}
We consider a spin-1/2 Heisenberg XXZ chain in the presence of a gradient magnetic field,
\begin{eqnarray} \label{Hamiltonian}
\hat{H}=\sum_{l=-L}^L\big(\frac{J}{2}\hat{S}^+_l\hat{S}^-_{l+1}+h.c. +\Delta\hat{S}^z_l\hat{S}^z_{l+1}+lB\hat{S}^z_l\big).
\end{eqnarray}
Here, $\hat{S}^i_l(i=x,y,z)$ are spin-1/2 operators for the $l$-th site ($l=-L,...,0,...,L$),
$\hat{S}^{\pm}_l=\hat{S}^x_l\pm i\hat{S}^y_l$ are spin raising and lowing operators for the $l$-th site,
$J$ is the spin exchange energy which is set as unit (i.e., $J=\hbar=1$),
$\Delta$ is the interaction between nearest-neighbor spins,
and $B$ is the magnetic field gradient.

When $B=0$, there are three types of ground states: the critical phase in $-1<\Delta <1$, the ferromagnetic phase in $\Delta<-1$, and the anti-ferromagnetic phase in $\Delta>1$.
In ferromagnetic phase regime $\Delta < -1$ , the excitation of a magnon is equal to the flipping of one spin.
For a sufficiently large ferromagnetic interaction $\Delta \ll 0$, the ground state is a completely ferromagnetic state $\left|\mathbf{0}\right\rangle$ with all spins downward $\left|\downarrow\downarrow\downarrow...\downarrow\right\rangle$
or upward $\left|\uparrow\uparrow\uparrow...\uparrow\right\rangle$.
The multi-magnon excitations can be prepared by flipping spins in the completely ferromagnetic state.

The Hamiltonian~\eqref{Hamiltonian} exhibits a U(1) symmetry under global spin rotations around the $z$-axis and the number of its total spin $\hat{S}^z=\sum_{l}\hat{S}^z_l$ is conserved (i.e., $[\hat{H},\hat{S}^z]=0$).
This means that the subspaces with different numbers of magnon excitations are decoupled.
Using the mapping: $\{\left|\downarrow\right\rangle\leftrightarrow \left|0\right\rangle, \left|\uparrow\right\rangle\leftrightarrow \left|1\right\rangle, \hat{S}^+_l\leftrightarrow\hat{a}_{l}^{\dag}, \hat{S}^-_l\leftrightarrow\hat{a}_l, \hat{S}^z_l\leftrightarrow\hat{n}_l-\frac{1}{2}\}$,
the Hamiltonian~\eqref{Hamiltonian} can be mapped onto
$\hat{H}=\sum\limits_{l} \left(\frac{J}{2}\hat{a}_{l}^{\dag}\hat{a}_{l+1}+h.c. +\Delta\hat{n}_l\hat{n}_{l+1}+B l\hat{n}_l\right)$,
with $\hat{n}_l=\hat{a}_{l}^{\dag}\hat{a}_l$.
Here, $\hat{a}_{l}^{\dag}$ ($\hat{a}_l$) is particle creation (annihilation) operator at the $l$-th site and they satisfy the commutation relations of hard-core bosons.
Thus one can understand the magnon excitations in the picture of hard-core bosonic system.
%

Inter-magnon interaction is a key ingredient in the formation of BSs and also has an influence on the dynamics.
Intuitively, ferromagnetic interactions ($\Delta<0$) and anti-ferromagnetic ones ($\Delta>0$) may affect the dynamics in different ways.
Nevertheless, a symmetry protected dynamical symmetry (SPDS) theorem~\cite{JYu2017} reveals a symmetric relation of the time-evolution observable between the repulsive and attractive systems, i.e., dynamical symmetry.
Combining with the time-reversal operator $\hat{R}$ and a unitary operator $\hat{W}$, an anti-unitary operator $\hat{Q}$ is defined as $\hat{Q}=\hat{R}\hat{W}$.
The SPDS theorem indicates that if the system follows three conditions:
\begin{itemize}
\item $\{\hat{Q},\hat{H}'\}=0,[\hat{Q},\hat{H}'']=0$ for the two parts of the Hamiltonian $\hat{H}=\hat{H}'+\hat{H}''$,
\item $\hat{Q}^{-1}|\psi(0)\rangle= e^{i \theta}|\psi(0)\rangle$ for the initial state $|\psi(0)\rangle$ and $\theta$ is a global phase factor, and
\item $\hat{Q}^{-1} \hat{O} \hat{Q}=\pm \hat{O}$ for a certain observable $\hat{O}$,
\end{itemize}
the system is able to manifest an interaction-induced dynamical symmetry for the time evolution of a certain observable.

According to the SPDS theorem, we divide the Hamiltonian~\eqref{Hamiltonian} into two parts $\hat{H}=\hat{H}'+\hat{H}''$ with
\begin{eqnarray}
\hat{H}'=\frac{1}{2}\sum_{l}\big(\hat{S}^+_l\hat{S}^-_{l+1}+\hat{S}^-_l\hat{S}^+_{l+1}\big),
\end{eqnarray}
and
\begin{eqnarray}
\hat{H}''=\Delta\sum_{l}\hat{S}^z_l\hat{S}^z_{l+1}+\sum_{l}lB\hat{S}^z_l.
\end{eqnarray}
For the system with only nearest-neighbor spin exchange, one can decompose it as odd-lattice $\mathcal{A}$ and even-lattice $\mathcal{B}$, thus we can define an operator $\hat{W}$ related to the bipartite lattice symmetry,
\begin{eqnarray}
\hat{W}^{-1}\hat{S}^-_l\hat{W}=\left\{
\begin{aligned}
\hat{S}^-_l,\ \text{if}\ l\in \mathcal{A}, \\
-\hat{S}^-_l,\ \text{if}\ l\in \mathcal{B}.
\end{aligned}
\right.
\end{eqnarray}
For the time-reversal operator $\hat{R}$, $\hat{R}^{-1}i\hat{R}=-i$, it is easy to conclude
\begin{equation}
\begin{array}{l}{\hat{R}^{-1} \hat{S}_{l}^{+} \hat{S}_{l+1}^{-} \hat{R}=\hat{R}^{-1}\left(\hat{S}_{l}^{x}+i \hat{S}_{l}^{y}\right)\left(\hat{S}_{l+1}^{x}-i \hat{S}_{l+1}^{y}\right) \hat{R}} \\ {=\hat{S}_{l}^{+} \hat{S}_{l+1}^{-}}\end{array}
\end{equation}
and
\begin{equation}
\hat{R}^{-1} \hat{S}_{l}^{z} \hat{S}_{l+1}^{z} \hat{R}=\hat{S}_{l}^{z} \hat{S}_{l+1}^{z}.
\end{equation}
Thus the $\hat{H}'$ and $\hat{H}''$ respectively satisfy the relations
\begin{equation}
\begin{array}{l}{\hat{Q}^{-1} \hat{H}' \hat{Q}=\frac{1}{2} \sum\limits_{l} \hat{W}^{-1} \hat{R}^{-1}\left(\hat{S}_{l}^{+} \hat{S}_{l+1}^{-}+\hat{S}_{l}^{-} \hat{S}_{l+1}^{+}\right) \hat{R} \hat{W}} \\ {=\frac{1}{2} \sum\limits_{l}\left(\hat{W}^{-1} \hat{S}_{l}^{+} \hat{W} \hat{W}^{-1} \hat{S}_{l+1}^{-} \hat{W}+\hat{W}^{-1} \hat{S}_{l}^{-} \hat{W} \hat{W}^{-1} \hat{S}_{l+1}^{+} \hat{W}\right)} \\ {=-\hat{H}'}\end{array}
\end{equation}
and
\begin{equation}
\begin{array}{l}
{\hat{Q}^{-1} \hat{H}'' \hat{Q}=\sum\limits_{l} \hat{W}^{-1}\left(\Delta \hat{S}_{l}^{z} \hat{S}_{l+1}^{z}+l B \hat{S}_{l}^{z}\right) \hat{W}} \\ =\sum\limits_{l} \hat{W}^{-1} \left(4\Delta[\hat{S}_{l}^{+}, \hat{S}_{l}^{-}]\cdot[\hat{S}_{l+1}^{+}, \hat{S}_{l+1}^{-}]+2lB[\hat{S}_{l}^{+}, \hat{S}_{l}^{-}]\right) \hat{W} \\
{=\hat{H}''}
\end{array}
\end{equation}
with $\hat{S}_{l}^{z}=2[\hat{S}_{l}^{+}, \hat{S}_{l}^{-}]$.
This anti-unitary operator $\hat{Q}$ ensures that $\hat{Q}$ anti-commutes with $\hat H'$ and commutes with $\hat H''$,
\begin{eqnarray}
\{\hat{Q},\hat{H}'\}=0, ~[\hat{Q},\hat{H}'']=0. \label{Con1}
\end{eqnarray}
Below we concentrate on discussing the time-evolution from the initial state of two-magnon excitations over the fully ferromagnetic state, $\left|\psi(0)\right\rangle=\hat{S}^+_{l_1^i}\hat{S}^+_{l_2^i} \left|\mathbf{0}\right\rangle$, where
$\left|\mathbf{0}\right\rangle = \left|\downarrow\downarrow\downarrow ...\downarrow\right\rangle$ and $l_1^i\neq l_2^i$.
Obviously, the initial state is invariant under the transformation $\hat{Q}$,
\begin{equation}
\hat{Q}^{-1}\left|\psi(0)\right\rangle =-\left|\psi(0)\right\rangle, \label{Con2}
\end{equation}
but there appears a global phase factor.
The spin correlations $\hat{S}^z_{l'}\hat{S}^z_{l''}$ between sites $l'$ and $l''$ satisfy
\begin{eqnarray}
\hat{Q}^{-1}\hat{S}^z_{l'}\hat{S}^z_{l''}\hat{Q} =\hat{S}^z_{l'}\hat{S}^z_{l''}. \label{Con3}
\end{eqnarray}
Defining
\begin{eqnarray}
C_{l',l''}(t)=\langle\psi(t)|\hat{S}^z_{l'}\hat{S}^z_{l''} |\psi(t)\rangle,
\end{eqnarray}
from Eqs.~\eqref{Con1}, \eqref{Con2}, \eqref{Con3} and $\hat{Q}^{-1}e^{-iHt}\hat{Q}=e^{\hat{Q}^{-1}({-iHt})\hat{Q}}$, we obtain,
\begin{eqnarray}\nn
&&C_{l',l''}(t)_{(\Delta,B)} =\langle\psi(0)|e^{i(\hat{H}'+\hat{H}'')t}\hat{S}^z_{l'}\hat{S}^z_{l''} e^{-i(\hat{H}'+\hat{H}'')t}|\psi(0)\rangle \\ \nn
&&=\langle\psi(0)|\hat{Q}e^{i(\hat{H}'+\hat{H}'')t}\hat{Q}^{-1} \hat{S}^z_{l'}\hat{S}^z_{l''}\hat{Q}e^{-i(\hat{H}'+\hat{H}'')t} \hat{Q}^{-1}|\psi(0)\rangle \\ \nn
&&=\langle\psi(0)|e^{i(\hat{H}'-\hat{H}'')t}\hat{S}^z_{l'}\hat{S}^z_{l''} e^{-i(\hat{H}'-\hat{H}'')t}|\psi(0)\rangle \\
&&=C_{l',l''}(t)_{(-\Delta,-B)}.\label{Con4}
\end{eqnarray}
This means that the time-dependent spin correlation is the same when we simultaneously change the signs of interaction $\Delta$ and magnetic field gradient $B$.
We can find a direct connection between the spin correlations with ($\Delta$,$B$) and ($-\Delta$,$-B$).
Different from the single-point operators in~\cite{JYu2017}, we extend its conclusions to a two-point operator.

When the sign of gradient magnetic field is flipped, the system~\eqref{Hamiltonian} is equivalent if we reverse the lattice around the centroid position of the initial state $l_c^i=\frac{l_1^i+l_2^i}{2}$, this is, the lattice index is changed from $l$ to $2 l_c^i-l$.
Then we have
\begin{eqnarray}
C_{l',l''}(t)_{(-\Delta,-B)}= C_{2l_c^i-l',2l_c^i-l''}(t)_{(-\Delta,B)}.\label{Con5}
\end{eqnarray}
Combining \eqref{Con4} and \eqref{Con5}, we can conclude
\begin{eqnarray}
C_{l',l''}(t)_{(\Delta,B)}= C_{2l_c^i-l',2l_c^i-l''}(t)_{(-\Delta,B)}.\label{Con6}
\end{eqnarray}
The relation \eqref{Con6} indicates that, when the sign of interaction $\Delta$ is changed, the time-evolution of spin correlation is symmetrical about the centroid position $l^i_{c}$ of the initial state.
This is to say, once we know the spin correlation under ferromagnetic interactions, we can deduce the results under anti-ferromagnetic interactions.
Therefore, below we only consider the system with ferromagnetic interactions $\Delta< 0$.

\begin{figure}[htp]
\begin{center}
\includegraphics[width=\columnwidth]{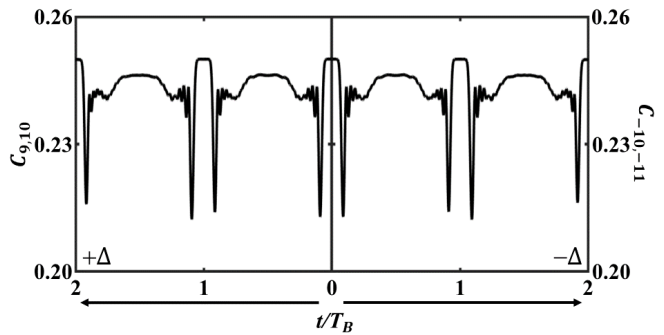}
\end{center}
\caption{(Color online) Dynamical symmetry of spin correlations under the parameters $(\Delta,B)$ and $(-\Delta,B)$. The time-evolution of two-spin correlations in 9th and 10th with ferromagnetic interaction (left half) and the ones in -10th and -11th with anti-ferromagnetic interaction (right half).}
\label{fig:1}
\end{figure}

In Fig.~\ref{fig:1}, starting from the initial state $\left|\psi(0)\right\rangle=\hat{S}^+_{-1}\hat{S}^+_{0} \left|\mathbf{0}\right\rangle$, we compare the spin correlation $C_{-10,-11}(t)$ with anti-ferromagnetic interaction ($-\Delta$) and the spin correlation $C_{9,10}(t)$ with ferromagnetic interaction ($\Delta$).
The parameters are chosen as $\Delta=-1.5$, $B=0.05$ and the total chain length $L_t=101$.
The numerical results completely follow the relation~\eqref{Con6}.

\section{signature of Bloch oscillations in spin correlations} \label{Sec3}

In this section, we consider how the interactions affect the dynamics of spin excitations under a gradient magnetic field.
Here the initial state is chosen as $|\downarrow...\downarrow\uparrow\uparrow\downarrow... \downarrow\rangle$, which means two-magnon excitations at adjacent sites.
In the time-evolution, the spin excitations will undergo BOs.
To show the interaction effects on BOs, we calculate the spin distributions
\begin{equation}
S^z_{l}(t)=\langle\psi(t)|\hat{S}^z_{l}|\psi(t)\rangle
\end{equation}
as a function of time, and the instantaneous longitudinal spin-spin correlations $C_{l',l''}(t)$ between sites $l'$ and $l''$ among the spin chain.
In the absence of interaction, we recover the results of traditional individual-particle BOs with frequency $\omega_B=B$.
Under strong interaction, we find a doubled Bloch frequency $\omega^{eff}_{B}=2B$, which indicates the appearance of interaction-induced fractional BOs of magnon BSs.
In the moderate-interaction case, we find the coexistence of individual-particle and bound-state BOs.

\begin{figure}[htp]
\center
\includegraphics[width=0.45\textwidth]{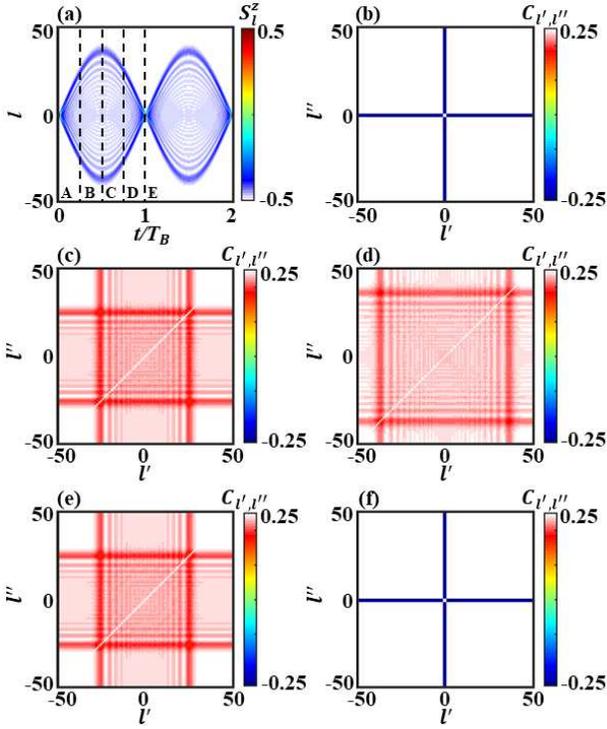}
\caption{(Color online) The Bloch dynamics of free magnons. (a) The spin distributions $S^z_{l}$ versus the rescaled time $t/T_B$.
(b)-(f) The longitudinal spin-spin correlations $C_{l',l''}$
at the different time marked as A-E in (a).
The parameters are chosen as $\Delta=0$, $B=0.05$, and the total chain length $L_t=101$.}
\label{fig:2}
\end{figure}

\subsection{Dynamics of free magnons} \label{SIIIA}

We first consider the BOs of magnons in a noninteracting system.
The eigen-values form a Wannier-Zeeman ladder with equidistant level spacing $\Delta E=B$ ~\cite{Yke2015}, which directly gives the Bloch frequency.
We numerically compute the time-evolution of the longitudinal spin-spin correlations via the TEBD algorithm.
The parameters are chosen as $\Delta=0$, $B=0.05$, and the total chain length $L_t=101$.
The noninteracting magnons independently undergo BOs with period $T_B=\frac{2 \pi}{B}$, see Fig.~\ref{fig:2} (a).
The free magnons periodically widen and shrink in an interval~\cite{THartmann2004}
\begin{equation}
|l|<\frac{2}{B}|\sin\frac{Bt}{2}|=l_b.
\end{equation}
Since there is no interaction, each magnon behaves as a free particle and the time-evolution recovers the breathing mode in the single-particle BOs.

Now we discuss the connection between the longitudinal spin-spin correlations $C_{l',l''}$ and the two-magnon correlations  $\Gamma_{l',l''}(t) =\langle\psi(t)|\hat{S}^+_{l'}\hat{S}^+_{l''} \hat{S}^-_{l''}\hat{S}^-_{l'} |\psi(t)\rangle$.
The two-magnon correlations $\Gamma_{l',l''}$ describe the probability of detecting one magnon at site $l'$ and the other one at site $l''$.
From their definitions, one can find
\begin{eqnarray}
C_{l',l''}=\Gamma_{l',l''} -\frac{1}{2}S^z_{l'} -\frac{1}{2}S^z_{l''}-\frac{1}{4},
\label{Con7}
\end{eqnarray}
with $l'\ne l''$, see Appendix~\ref{appendixC} for more details.
Due to $(\hat{S}^z_{l})^2=0.25$, the longitudinal spin-spin correlations $C_{l',l''}$ are always equal to 0.25 for $l'=l''$.
Therefore we set ${C_{l',l''}}=0.25$ as the background value of the longitudinal spin-spin correlations.
As every site in the region of $|l|>l_b$ is spin down and $\Gamma_{l',l''}=0$ in the region of $|l'|>l_b$ or $|l''|>l_b$, the relation~\eqref{Con7} is further given as
\begin{eqnarray}
C_{l',l''}=\left\{
\begin{aligned}
-\frac{1}{2}S^z_{l'},\ \text{if}\ |l''|>l_b, l'\ne l'', \\
-\frac{1}{2}S^z_{l''},\ \text{if}\ |l'|>l_b, l''\ne l'.
\end{aligned}
\right.
\label{Con8}
\end{eqnarray}
Thus, (i) in the region of $|l'|<l_b$ and $|l''|>l_b$, $C_{l',l''}$ show fringes and may not equal to the background value 0.25; and (ii) in the region of $|l'|>l_b$ and $|l''|>l_b$, $C_{l',l''}=\langle\psi(t)|\hat{S}^z_{l'}\hat{S}^z_{l''} |\psi(t)\rangle=(-0.5)\times(-0.5)=0.25$.

Different from $C_{l',l''}$, only in the region of $(-l_b,l_b)$, $\Gamma_{l',l''}$ are nonzero.
Obviously, the cross-overlapping region in $C_{l',l''}$ is well consistent with the region of nonzero $\Gamma_{l',l''}$.
The significant $\Gamma_{l',l'+1}$ correspond to the BOs of bounded magnons.
If there is no significant $\Gamma_{l',l'+1}$, the BOs of free magnons dominate.

For an example, in Fig.~\ref{fig:2}~(d), when the spin excitations propagate to $l_b=40$ at $t_b=T_B/2$, in the region of $|l|<40$ we have $-0.5\leq S^z_{l}(t)\leq 0.5$, while in the region of $|l|>40$ we have $S^z_{l}(t)=-0.5$.
From the relation~\eqref{Con8}, in the four corners of $|l'|>40$ and $|l''|>40$, $C_{l',l''}=\langle\hat{S}^z_{l'}\hat{S}^z_{l''} \rangle=(-0.5)\times(-0.5)=0.25$.
Moreover, the longitudinal spin-spin correlations satisfy $-0.25\leq C_{|l'|< 40,|l''|>40}\le 0.25$ and $-0.25\leq C_{|l'|> 40,|l''|< 40}\le 0.25$ and show fringes.
The basic pattern of the longitudinal spin-spin correlations behaves like a cross in the $(l',l'')$ plane.
The cross-overlapping region in $C_{l',l''}$ is well consistent with the region of nonzero $\Gamma_{l',l''}$, see Fig.~\ref{fig:2}~(d) and the top panel in the third column of Fig.~\ref{fig:8}.
The cross-overlapping region in $C_{l',l''}$ is determined by the amplitude of BOs, see the point C in Fig.~\ref{fig:2}~(a).
These numerical results are well consistent with the analytical ones~\eqref{Con7} and~\eqref{Con8}.

In Figs.~\ref{fig:2} (b)-(f), we show the $C_{l',l''}$ at the different time: A ($t=0$), B ($t=T_B/4$), C ($t=T_B/2$), D ($t=3T_B/4$) and E ($t=T_B$).
%
%
%
As the spin excitations expand and shrink, the cross-overlapping region becomes larger at B ($t=T_B/4$), reaches the maximum at C ($t=T_B/2$), gradually decreases at D ($t=3T_B/4$) and finally
recovers the initial state at E ($t=T_B$).
The cross-overlapping regions in Figs.~\ref{fig:2} (b)-(f) are in excellent agreement with the regions of spin excitations in (A)-(E) in Fig.~\ref{fig:2} (a), respectively.

\subsection{Dynamics of strongly interacting magnons} \label{SIIIB}

Under strong interactions, through implementing the many-body degenerate perturbation analysis, we derive an effective single-particle Hamiltonian and explore the interaction-induced fractional BOs~\cite{MTakahashi1977,Lee2004,SBravyi2011,XQin2014}.

\begin{figure}[htp]
\center
\includegraphics[width=0.45\textwidth]{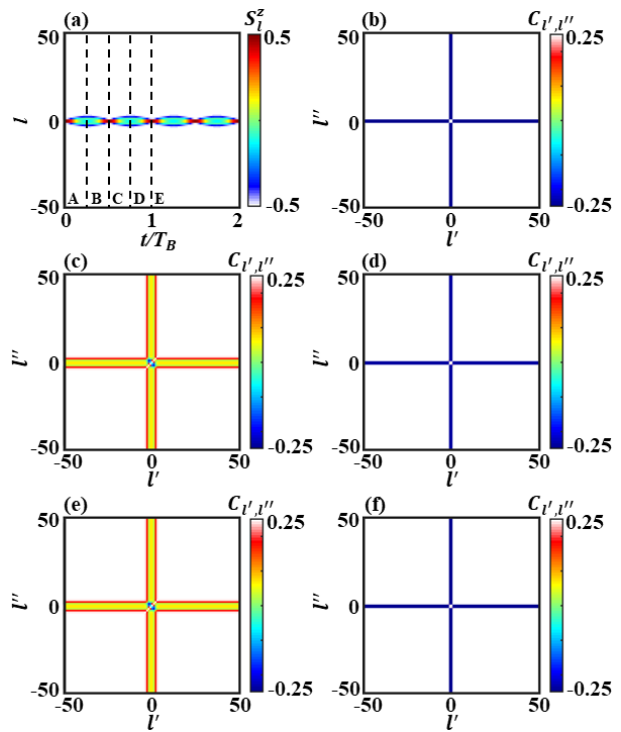}
\caption{(Color online) The Bloch dynamics of two strongly interacting magnons. (a) The spin distributions $S^z_{l}$ versus the rescaled time $t/T_B$.
(b)-(f) The longitudinal spin-spin correlations $C_{l',l''}$
at the different time marked as A-E in (a).
The parameters are the same as those in Fig.~\ref{fig:2} except for $\Delta=-5$.}
\label{fig:3}
\end{figure}

Under the condition of $|\Delta|\gg(1/2,|B|)$, one can treat the hopping term and the gradient magnetic field term
\begin{equation}
\hat{H}_1=\frac{1}{2}\sum_{l} \big(\hat{S}^+_l\hat{S}^-_{l+1} +\hat{S}^-_l\hat{S}^+_{l+1}\big)+\sum_{l} lB\hat{S}^z_l,
\end{equation}
as a perturbation to the interaction term
\begin{equation}
\hat{H}_0=\Delta\sum_{l}\hat{S}^z_l\hat{S}^z_{l+1}.
\end{equation}
The effective single-particle Hamiltonian can be written as
\begin{equation}
\hat{H}_{eff}=\frac{1}{4\Delta}\sum_{m}(\hat{C}_{m}^{\dag}\hat{C}_{m+1} +\hat{C}_{m+1}^{\dag}\hat{C}_m)
+\sum_{m}2Bm\hat{C}_{m}^{\dag}\hat{C}_m
\end{equation}
with $m=-L,...,0,...,L$,
whose detailed derivation can be found in Appendix~\ref{appendixA}.
Here, the operator $\hat{C}_{m}^{\dag}=\hat S_m^{+}\hat S_{m+1}^{+}$ means simultaneously flipping two adjacent spins at $m$-th and $(m+1)$-th sites.
The two-magnon excitations behave like a single particle in the tilted lattices with doubled frequency $\omega^{eff}_{B}=2B$.
The two magnons tend to travel together and undergo fractional BOs, see Fig.~\ref{fig:3}~(a).
The initial state and parameters are the same as those in Fig.~\ref{fig:2}, except for $\Delta=-5$.
The region of spin excitations is given as
\begin{equation}
|l|<\frac{1}{2\Delta B}|\sin\left(B t\right)|.
\label{Con9}
\end{equation}
Compared with Fig.~\ref{fig:2}~(a), the width is reduced by a factor $1/(4\Delta)$ while the oscillation frequency becomes double.
The numerical results are well consistent with the analytical ones~\eqref{Con9}.

$C_{l',l''}$ reveal the effective single-particle dynamics.
Similarly, in one period, $C_{l',l''}$ behave like a cross for each instantaneous state, see Figs.~\ref{fig:3}~(b)-(f).
As the spin excitations expand and shrink, the cross-overlapping region reaches the maximum at B ($t=T_B/4$), recovers the initial state at C ($t=T_B/2$), increases to the maximum again at D ($t=3T_B/4$) and finally recovers the initial state again at E ($t=T_B$).
It clearly manifests that the two strongly interacting magnons undergo a breathing motion with the half-period of the free-magnon breathing motion.
The cross-overlapping regions in Figs.~\ref{fig:3} (b)-(f) are in excellent agreement with the regions of spin excitations in (A)-(E) in Fig.~\ref{fig:3}~(a), respectively.

\subsection{Dynamics of moderately interacting magnons} \label{SIIIC}

\begin{figure}[htp]
\center
\includegraphics[width=0.45\textwidth]{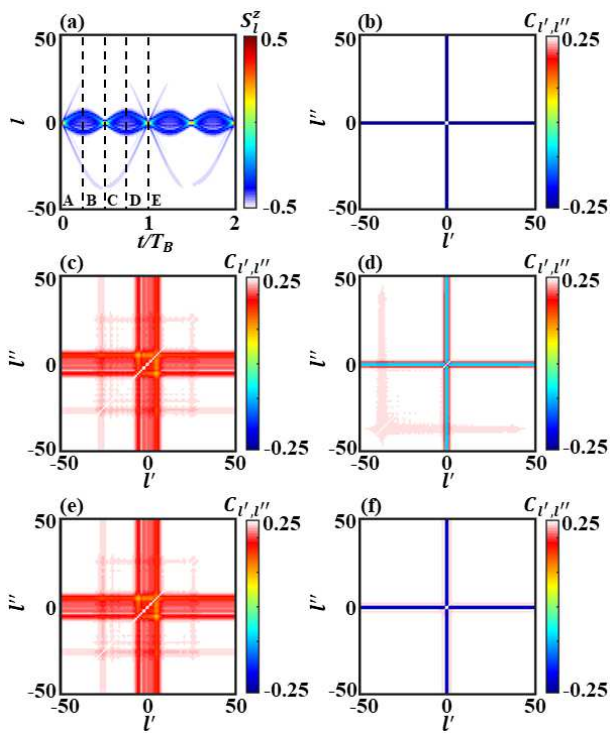}
\caption{(Color online) The Bloch dynamics of two moderately interacting magnons. (a) The spin distributions $S^z_{l}$ versus the rescaled time $t/T_B$.
(b)-(f) The longitudinal spin-spin correlations $C_{l',l''}$ at the different time marked as A-E in (a).
The parameters are the same as those in Fig.~\ref{fig:2} except for $\Delta=-1.5$. }
\label{fig:4}
\end{figure}

At last, we study the dynamics of two-magnon excitations in the  moderate-interaction case.
The initial state and the parameters are the same as those in Fig.~\ref{fig:2} except for $\Delta=-1.5$.
The spin distributions $S^z_{l}$ exhibit the coexistence of two breathing modes, see Fig.~\ref{fig:4}~(a).
The outer and inner breathing modes correspond to the oscillations of free magnons and bounded magnons, respectively.
This is because that the initial state is prepared as the superposition of scattering and bound states (see Appendix~\ref{appendixB} for more details).
Nevertheless, the outer breathing mode is slightly asymmetric about the initial position $l_c^i$, different from the breathing mode of free magnons.
The asymmetry may come from the interaction-induced scattering of free-magnon component.

Similar to the spin distributions $S^z_{l}$, $C_{l',l''}$ also show the coexistence of inner and outer pattern, see Figs.~\ref{fig:4}~(c)-(f) for different time: B ($t=T_B/4$), C ($t=T_B/2$), D ($t=3T_B/4$) and E ($t=T_B$).
The correlations partially recover the initial correlations at C ($t=T_B/2$).
This is because that the bound-state component returns to the initial ones while the scattering-state component is not yet.
At E ($t=T_B$), both two components nearly return to the initial state, see Fig.~\ref{fig:4}~(f).

So far, we examine the role of spin-spin interaction on the dynamics of magnon excitations among the spin chain.
By increasing the interaction strength, one may observe clear enhancement of the correlated tunneling of two magnons.
Moreover, the longitudinal spin-spin correlations can be utilized to characterize the multi-magnon BOs.

\section{Extracting magnetic field gradient from multi-magnon Bloch oscillations} \label{Sec4}

In this section, we discuss how to determine the magnetic field gradient from the multi-magnon BOs.
For convenience, we flip two neighboring spins in the completely ferromagnetic state  $\left|\downarrow\downarrow\downarrow ...\downarrow\right\rangle$.
When two-magnon excitations are launched on the adjacent sites of the spin chain under a gradient magnetic field, it may exhibit a dynamical localization in a period.
The time-dependent spin distributions and longitudinal spin-spin correlations both show the coexistence of two components when the interaction strength is moderate.
However, we cannot accurately determine the magnetic field gradient (which determines the Bloch frequency) via the spin distributions or longitudinal spin-spin correlations.
Below, we analyze the fidelity and the sub-standard deviation
in both time and frequency domains to extract the gradient of magnetic field and the Bloch frequency.

\subsection{Fidelity}

By simulating the time-evolution with the TEBD algorithm, we calculate the time-dependent fidelity
\begin{eqnarray}
F(t)=|\langle\Psi(0)|\Psi(t)\rangle|^2.
\end{eqnarray}
It characterizes the probability of the time-evolved state returning to the initial state.

\begin{figure}[htp]
\center
\includegraphics[width=0.45\textwidth]{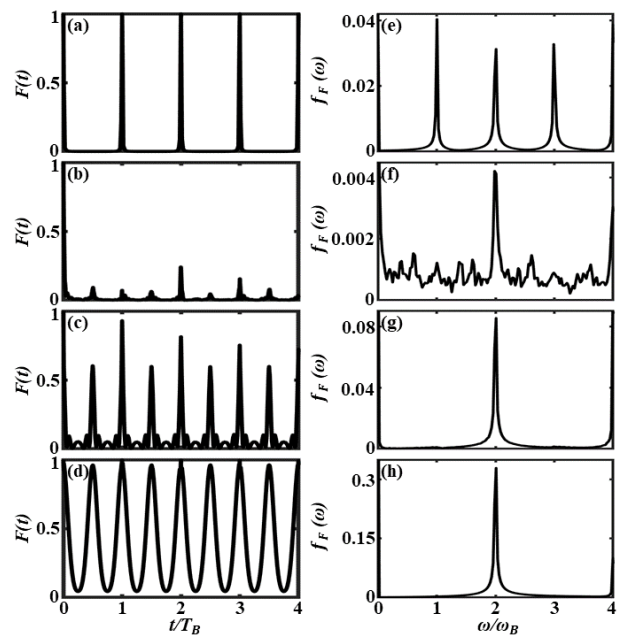}
\caption{(Color online) Left column: the fidelity $F$ versus the rescaled time $t/T_B$ for different values of $\Delta$: (a) $0$, (b) $-1$, (c) $-1.5$ and (d) $-5$.
Right column: the frequency distribution $f_F(\omega)$ of the fidelity for different values of $\Delta$: (e) $0$, (f) $-1$, (g) $-1.5$ and (h) $-5$.
The other parameters are chosen as $B=0.05$ and the total chain length $L_t=101$.}
\label{fig:5}
\end{figure}

A slight change of the interaction may have a huge influence on the dynamics.
We discuss the fidelity versus the rescaled time $t/T_B$ for different interaction strengths $\Delta$: (a) $0$, (b) $-1$, (c) $-1.5$ and (d) $-5$, see Fig.~\ref{fig:5}.
The gradient of magnetic field is chosen as $B=0.05$, and the total length of spin chain is $L_t=101$.
For clear visibility, the evolved time is set to be $t=4T_B$.
Without interaction, the sharp peaks perfectly emerge at the integer multiples of period $T_B$, see Fig.~\ref{fig:5}~(a).
When the interaction increases, in addition to the peaks at the integer multiples of period $T_B$, there also appear peaks at the integer multiples of half-period $T_B/2$, see Figs.~\ref{fig:5}~(b)-(d).
For the moderate-interaction strength $\Delta=-1.5$, we find the coexistence of peaks at both the integer multiples of periods $T_B/2$ and $T_B$, see Fig.~\ref{fig:5}~(c).
The period $T_B$ and half-period $T_B/2$ respectively correspond the free-magnon Bloch frequency $B$ and the bounded-magnon Bloch frequency $2B$.
For stronger interaction strength $\Delta=-5$ , the dynamics transfers from the independent BOs to the effective single-magnon BOs, and the half-period oscillation of fidelity is dominant, see Fig.~\ref{fig:5}~(d).
To explain how the interaction affects the fidelity, we project the initial state onto the scattering and the bound states.
We find that the occupation on BSs becomes larger as the interaction increases (see Appendix~\ref{appendixB} for more details).
Thus the peaks of fidelity at the half-period $T_B/2$ become higher as the interaction increases.

However, when the free-magnon component dominates in the weak-interaction case, it is difficult to distinguish the bound-state component from the scattering-magnon component by directly observing the time-evolution of the fidelity.
Under such a moderate interaction, the periodicity of fidelity is destroyed due to the appearance of irregular behaviors (such as quantum chaos), see Fig.~\ref{fig:5}~(b).

In Figs.~\ref{fig:5}~(e)-(h), we show the frequency distribution $f_F(\omega)$ of the fidelity $F(t)$ for different interaction strengths $\Delta$: (e) $0$, (f) $-1$, (g) $-1.5$ and (h) $-5$.
In the absence of interaction, the peaks $n\omega_B$ (with positive integers $n$) gradually decay, see Fig.~\ref{fig:5} (e).
Once the interaction is introduced, there appear significant peaks at $2\omega_B$ and the peaks at $\omega_B$ vanish, see Figs.~\ref{fig:5}~(f)-(h).
This means that, for moderate interaction strengths, $f_F(\omega)$ cannot successfully identify the free-magnon BOs and it is difficult to show the coexistence of free-magnon BOs and bounded-magnon fractional BOs.

\subsection{Sub-standard deviation}

In this subsection, we present how to use the sub-standard deviation to extract the multi-frequencies of BOs, especially when the free-magnon BOs and bounded-magnon fractional BOs coexist.
Here, we will analyze the frequency distribution of the time-dependent generalized standard deviation,
\begin{equation}
D^x(t)=\sqrt{\sum_{l}(\langle\hat{S}^z_l\rangle+1/2)|l-l_{c}(t)|^x},
\end{equation}
which can characterize the fluctuation of spin excitations in spatial distribution.
Here $\langle\hat{S}^z_l\rangle$ represents the spin magnetization at site ${l}$ and
\begin{equation}
{l_{c}(t)}=\frac{\sum_{l}l(\langle\hat{S}^z_l\rangle+1/2)}{\sum_{l}(\langle\hat{S}^z_l\rangle+1/2)}
\end{equation}
is the centroid position of the time-evolved state.
When $x=2$, the generalized standard deviation becomes traditional standard deviation.
Instead of traditional standard deviation, we define the super-standard deviation for $x>2$ and sub-standard deviation for $x<2$ to highlight the bounded- and free-magnon components, respectively.
After a series of trials, we find that one may choose sub-standard deviation with $x=1/2$ to extract the multiple Bloch frequencies.

\begin{figure}[htp]
\center
\includegraphics[width=0.45\textwidth]{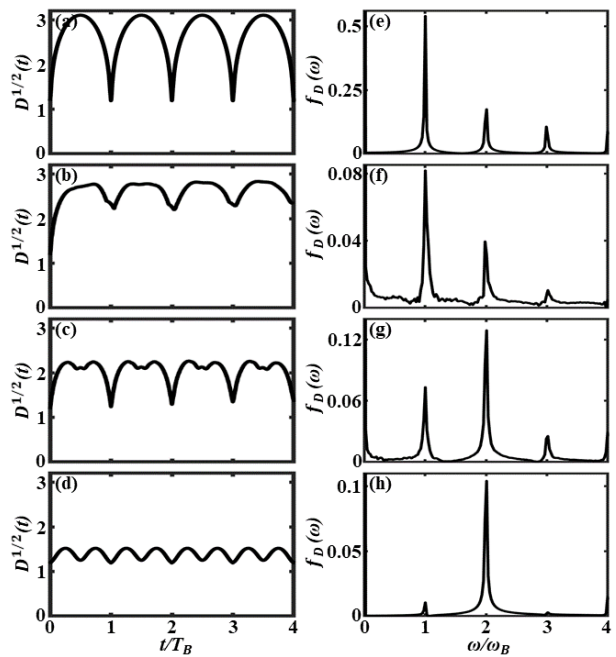}
\caption{(Color online) Left column: the sub-standard deviation $D^{1/2}$ versus the rescaled time $t/T_B$ for different values of $\Delta$: (a) $0$, (b) $-1$, (c) $-1.5$ and (d) $-5$.
Right column: the frequency distribution $f_D(\omega)$ of the sub-standard deviation for different values of $\Delta$: (e) $0$, (f) $-1$, (g) $-1.5$ and (h) $-5$.
The other parameters are chosen as $B=0.05$ and the total chain length $L_t=101$.}
\label{fig:6}
\end{figure}

The time-dependent sub-standard deviations show that the spatial region of spin excitations decreases as the interaction strength increases, see Figs.~\ref{fig:6}~(a)-(d).
Making a fast Fourier transform of the sub-standard deviations to obtain the frequency distribution $f_D(\omega)$,
one can observe sharp peaks centering at the integer multiples of $\omega_B$ in the frequency domain.
The maximum peak centers at $\omega_B$ for noninteracting systems, see Fig.~\ref{fig:6}~(e).
As the interaction increases, the peak at $\omega_B$ becomes lower while the peak at $2\omega_B$ becomes higher, and the peak at $2\omega_B$ becomes dominant under strong interactions, see Figs.~\ref{fig:6}~(f)-(h).
Since the peaks at $\omega_B$ and $2\omega_B$ are mainly induced by BOs of scattering and bounded magnons, respectively,
the dominant of peak at $2\omega_B$ is a clear signature of two-magnon BSs.

Unlike $f_F(\omega)$, the frequency distribution $f_D(\omega)$ of sub-standard deviation may clearly show the coexistence of the free-magnon BOs and bounded-magnon fractional BOs.
This means that $f_D(\omega)$ is a better quantity to witness the coexistence and competition between free-magnon BOs and bounded-magnon fractional BOs.
But for $f_F(\omega)$, although there appear significant peaks at $2\omega_B$, the peak at $\omega_B$ nearly vanish and thus cannot show the coexistence of free-magnon BOs and bounded-magnon fractional BOs.

There are two typical schemes to measure the force based on the delocalization-enhanced BOs and driving resonance tunneling effects~\cite{MGTarallo2012}.
Here we find that, once we determine the position of the peak $\omega_B$ or $2\omega_B$, the magnetic field gradient can be accurately given.
Moreover, bounded magnons undergo fractional BOs with frequency doubling $2\omega_B$.
The fractional BOs are an excellent indicator for judging the appearance of two-magnon BSs.

\section{Summary and Discussions} \label{Sec5}
In this work, through considering a Heisenberg XXZ chain under a gradient magnetic field, we study how the interaction affects the BOs of two-magnon excitations and give a quantitative method to extract the magnetic field gradient from the multi-frequency BOs.
We extend the theory of dynamical symmetry of single-point operators to the one of two-point operators, and find that the dynamics in anti-ferromagnetic systems can be directly derived from the corresponding ferromagnetic ones.
As the interaction increases, we find that the spin distribution or longitudinal spin-spin correlation dynamics gradually transfers from BOs of free magnons to the fractional BOs of bounded magnons.
The interaction-induced fractional BOs provide a new perspective to observe the magnon BSs.
Moreover, we use the fidelity and the sub-standard deviation in both time and frequency domains to probe the multi-frequency BOs and determine the magnetic field gradient.
The sub-standard deviation is an excellent candidate to witness the coexistence and competition between free-magnon BOs (at frequency $\omega_B$) and bounded-magnon fractional BOs (at frequency $2\omega_B$).

Based on the current techniques in engineering ultracold atoms, it is possible to simulate our Heisenberg spin chain.
By loading two-state $\rm ^{87}Rb$ atoms into a one-dimensional optical lattice in the Mott regime with one particle per lattice site,
the two hyperfine states with different magnetic dipole moments can be labeled as spin-up and spin-down, respectively.
Applying a gradient magnetic field along the lattice, our spin-dependent lattices can be realized.
The dynamics of spin distribution and longitudinal spin-spin correlation can be tracked via the techniques of atomic microscope~\cite{TFukuhara20131,TFukuhara20132}.
The interaction between magnons can be tuned via Feshbach resonance techniques~\cite{AWidera2004,CGross2010}.
With the observed spin distributions and longitudinal spin-spin correlations, the fidelity and the sub-standard deviation can be given.

\begin{acknowledgments}
This work was supported by the National Natural Science Foundation of China (Grants No. 11874434 and No. 11574405).
Y.K. was partially supported by the International Postdoctoral Exchange Fellowship Program (Grant No. 20180052).
\end{acknowledgments}

\appendix

\section{Effective single-particle Hamiltonian for strongly interacting magnons} \label{appendixA}

Under strong ferromagnetic interactions, the magnons prefer to travel together instead of the individual propagation.
In order to explain this phenomenon, we analytically construct an effective single-particle Hamiltonian by using the many-body degenerate perturbation theory.

When$|\Delta|\gg (1/2,|B|)$, we can divide the Hamiltonian into the $\hat{H}_0$ as a dominant term and $\hat{H}_1$ as a perturbation term.
In the two-magnon basis $\{|l'_1l'_2\rangle=\hat{S}^+_{l'_1}\hat{S}^+_{l'_2} |\textbf{0}\rangle:-L\leq l'_1<l'_2\leq L\}$, the $\hat{H}_0$ consists of two subspaces $\mathcal{U}$ and $\mathcal{V}$.
The total chain length $L_t=2L+1$. The degenerate eigen-states $\{|G_m\rangle=|m,m+1\rangle:-L\leq m\leq L\}$ form the subspace $\mathcal{U}$ with eigen-values $E_0=\Delta$.
Correspondingly, the degenerate eigen-states $\{|E_{l_1l_2}\rangle=|l_1l_2\rangle:l_1\neq l_2\pm 1,-L\leq l_1<l_2\leq L\}$ form the subspace $\mathcal{V}$ with eigen-values $E_1=0$.
The projection operators define as $\hat{P}_{\mathcal{U}}=\sum\limits_m |G_m\rangle\langle G_m|$ onto $\mathcal{U}$ and $\hat{P}_{\mathcal{V}}=\sum\limits_{l_2\neq l_1\pm 1}\frac{1}{E_0-E_1}|E_{l_1l_2}\rangle\langle E_{l_1l_2}|$ onto $\mathcal{V}$.
The second-order effective Hamiltonian is written as
\begin{eqnarray}\nn
\hat{H}_{eff}&=&\hat{h}_0+\hat{h}_1+\hat{h}_2 \\
&=&E_0\hat{P}_{\mathcal{U}}+\hat{P}_{\mathcal{U}}\hat{H}_1\hat{P}_{\mathcal{U}}+\hat{P}_{\mathcal{U}}\hat{H}_1\hat{P}_{\mathcal{V}}\hat{H}_1\hat{P}_{\mathcal{U}}.
\end{eqnarray}
The first-order perturbation reads as
\begin{eqnarray}\nn
\hat{h}_1&=&\hat{P}_{\mathcal{U}}\hat{H}_1\hat{P}_{\mathcal{U}} \\
&=&\sum\limits_{l,mm'} |G_m\rangle\langle G_m|(lB\hat{S}^z_l)|G_{m'}\rangle\langle G_{m'}|.
\end{eqnarray}
Since
\begin{eqnarray}\nn
&\sum\limits_{l}&\langle G_m|(lB\hat{S}^z_l)|G_{m'}\rangle \\ \nn
&=&B\delta_{mm'}\sum\limits_l l(\delta_{lm}+\delta_{l,m+1}-\frac{1}{2}) \\
&=&B\delta_{mm'}(2m+1),
\end{eqnarray}
we have
\begin{eqnarray}
\hat{h}_1=B\sum\limits_m(2m+1)|G_m\rangle\langle G_m|. \label{firstorder}
\end{eqnarray}
The second-order perturbation reads as
\begin{eqnarray}\nn
\hat{h}_2&=&\hat{P}_{\mathcal{U}}\hat{H}_1\hat{P}_{\mathcal{V}} \hat{H}_1\hat{P}_{\mathcal{U}} \\ \nn
&=&\frac{1}{4\Delta}\sum\limits_{mm',ll',l_1l_2} [|G_m\rangle\langle G_m|(\hat{S}^+_l\hat{S}^-_{l+1} +\hat{S}^-_l\hat{S}^+_{l+1}) |E_{l_1l_2}\rangle \\
&&\times\langle E_{l_1l_2}|(\hat{S}^+_l\hat{S}^-_{l+1} +\hat{S}^-_l\hat{S}^+_{l+1}) |G_{m'}\rangle\langle G_{m'}|].
\end{eqnarray}
After a careful calculation, we have
\begin{eqnarray}\nn
\hat{h}_2&=&\hat{P}_{\mathcal{U}}\hat{H}_1 \hat{P}_{\mathcal{V}}\hat{H}_1\hat{P}_{\mathcal{U}} \\
&=&\frac{1}{4\Delta}\sum\limits_{m}(|G_m\rangle+|G_{m+1}\rangle) (\langle G_m|+\langle G_{m+1}|).\label{secondorder}
\end{eqnarray}
Combing Eqs.~\eqref{firstorder} and \eqref{secondorder}, we derive the effective single-particle Hamiltonian up to the second order
\begin{eqnarray}\nn
\hat{H}_{eff}=&\displaystyle \frac{1}{4\Delta}&\sum\limits_{m}(|G_m\rangle+|G_{m+1}\rangle) (\langle G_m|+\langle G_{m+1}|) \\ \nn
&+&B\sum\limits_{m}(2m+1)|G_m\rangle\langle G_m| \\
&+&\Delta\sum\limits_{m}|G_m\rangle\langle G_m|
\end{eqnarray}
with $m=-L,...,0,...,L$.
We introduce the operator $\hat{C}_{m}^{\dag}=\hat{S}^+_m\hat{S}^+_{m+1}$, which means simultaneously flipping two adjacent spins at $m$-th and $(m+1)$-th sites from the vacuum state $|\mathbf{0}\rangle=|\downarrow\downarrow\ldots\downarrow\rangle$.
Therefore, the bound pairs behave as a composite particle following the Hamiltonian
\begin{eqnarray}\nn
\hat{H}_{eff}=&\displaystyle \frac{1}{4\Delta}&\sum\limits_{m}(\hat{C}_{m}^{\dag}\hat{C}_{m+1} +\hat{C}_{m+1}^{\dag}\hat{C}_{m}) \\
&+&\sum\limits_{m}2Bm\hat{C}_{m}^{\dag}\hat{C}_{m},
\end{eqnarray}
where the energy constant is omitted.

Compared with free magnons, the formation of bound pairs performs BOs with the doubled frequency.

\section{Two-magnon energy spectrum} \label{appendixB}

\begin{figure}[htp]
\center
\includegraphics[width=0.45\textwidth]{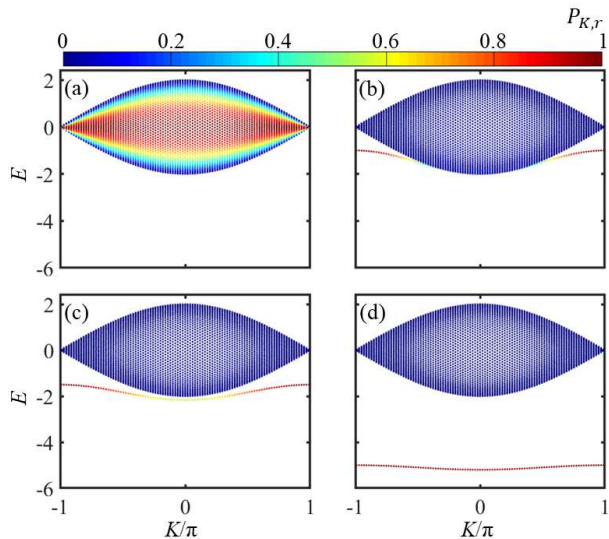}
\caption{(Color online) Two-magnon energy spectrum.
The two-magnon energy spectrum for $B=0$, $L_t=101$ and different values of $\Delta$: (a) $0$, (b) $-1$, (c) $-1.5$ and (d)
$-5$.
In which, the overlaps $P_{K,r}$ between the initial state with each eigen-states are denoted by the colorbar.}
\label{fig:7}
\end{figure}

To explain the interaction effects on magnon dynamics, we calculate the overlaps of initial state with the scattering and bound states in the absence of gradient magnetic field.

Since $[\hat{H},\hat{S}^z]=0$ with $\hat{S}^z=\sum_{l}\hat{S}^z_l$, the total number of spin excitations is conserved and all states keep evolving in the two-magnon Hilbert space.
The two-magnon Hilbert space is spanned by the basis $\mathcal{B}^{(2)}={\big{|l^{\prime}_1l^{\prime}_2\rangle =\hat{S}^+_{l^{\prime}_1}\hat{S}^+_{l^{\prime}_2} |\textbf{0}\rangle\big.}}$
with $-L\leq l'_1<l'_2\leq L$ and the total chain length $L_t=101$.
The eigen-states can be expressed as $|\Psi\rangle=\Sigma_{l'_1<l'_2}\Psi_{l'_1l'_2}|l'_1l'_2\rangle$ with $\Psi_{l'_1l'_2}=\langle\textbf{0} |\hat{S}^-_{l'_2}\hat{S}^-_{l'_1}|\Psi\rangle$.
Thus the system satisfies the following eigen-equation
\begin{eqnarray}
E\Psi_{l_1l_2}&=&\frac{1}{2}\big(\Psi_{l_1,l_2+1} +\Psi_{l_1,l_2-1}+\Psi_{l_1+1,l_2}+\Psi_{l_1-1,l_2}\big) \nonumber \\
&+&\Delta(\delta_{l_1,l_2-1}+\delta_{-L,L})\Psi_{l_1l_2}. \label{eigenequation}
\end{eqnarray}
In the absence of gradient magnetic field, the Heisenberg XXZ chain has a co-translational symmetry and the center-of-mass quasi-momentum is a good quantum number under the periodic boundary condition.
The motion of the two-magnon excitations consists of the motion of the center-of-mass $R=\frac{1}{2}(l_1+l_2)$ and the relative position $r=l_1-l_2$.
Defining $\Psi_{l_1l_2}=e^{iKR}\phi(r)$, the eigen-equation~\eqref{eigenequation} reads
\begin{equation}
E\phi(r)=\cos\frac{K}{2}\big(\phi(r-1)+\phi(r+1)\big) +\Delta\delta_{r,\pm 1}\phi(r).
\end{equation}
Under the periodic boundary conditions, we find $e^{iKL_t}=1$ and $\phi(r+L_t)=e^{iKL_t/2}\phi(r)$ with the quasi-momentum $K=2\pi\alpha/L_t$ (with $\alpha=-L,-L+1,...,L$).
Moreover, we have $\phi(0)=0$ and $\phi(r)=\phi(-r)$ with the commutation relations.

We give the two-magnon energy spectrum by numerically diagonalizing the Hamiltonian without a gradient magnetic field.
With the nearest-neighbor ferromagnetic interaction, the two magnons are able to form BSs.
When the interaction $|\Delta|>1$, the energy spectrum shows that BSs (corresponding to the lower band) completely separate from the scattering states (SSs).
After calculating the overlaps
\begin{eqnarray}
P_{K,r}=|\langle\psi_{K,r}|\psi(0)\rangle|^2,
\end{eqnarray}
we exactly reveal the proportion of initial state in each eigen-states $\psi_{K,r}$, see the color of energy spectrum in Fig.~\ref{fig:7}.
The interaction values $\Delta$ are set as (a) $0$, (b) $-1$, (c) $-1.5$ and (d)
$-5$.
The proportion in SSs makes the spins undergo independent BOs, while the proportion in BSs induces the correlated and fractional BOs.
Once the initial state strongly overlaps with BSs, one can clearly observe the signature in the time-evolution of spin distributions.
As the interaction increases, the overlaps with BSs become larger and the BOs of bounded magnons become dominant.

\section{Two-magnon correlations} \label{appendixC}

\begin{figure*}[htb]
\center
\includegraphics[width=1\textwidth]{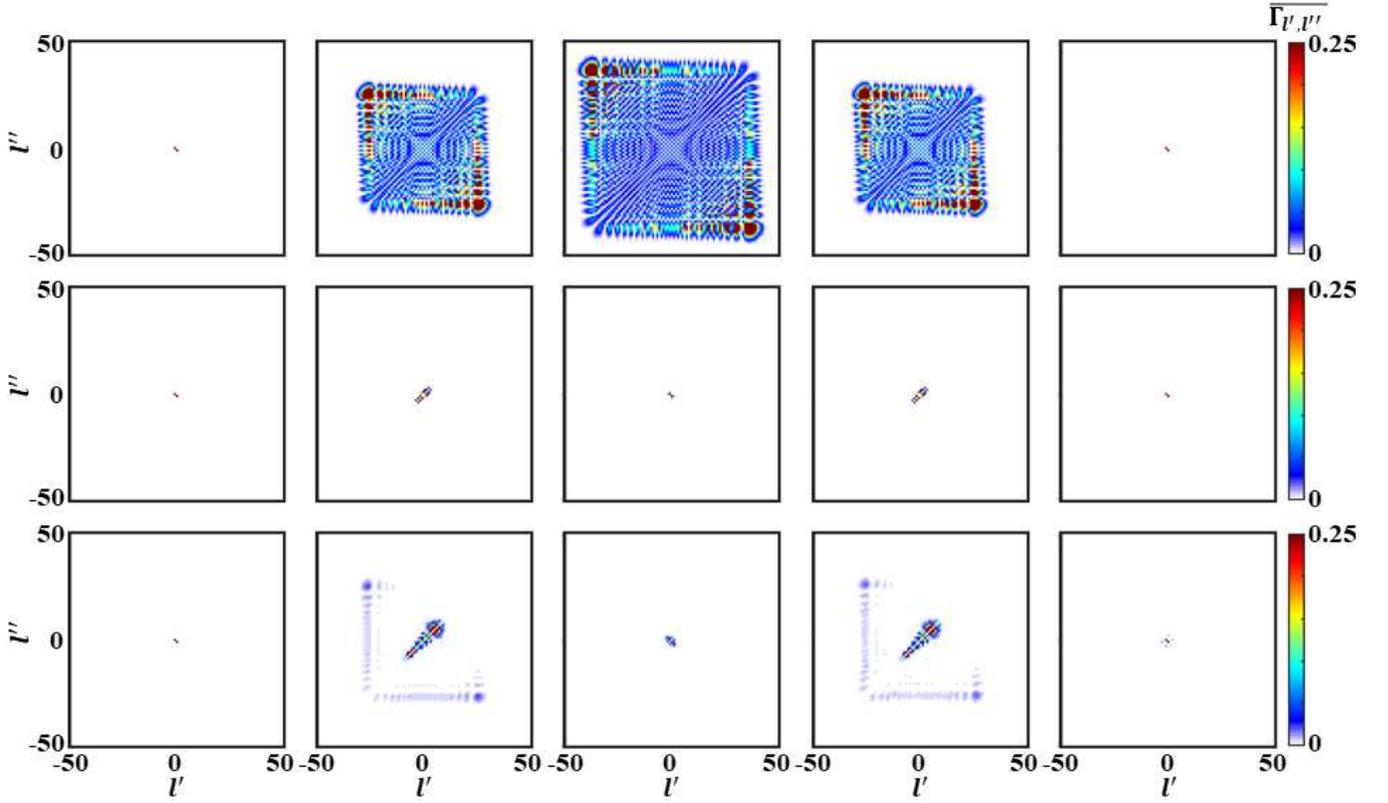}
\caption{(Color online) The rescaled two-magnon correlations $\overline{\Gamma_{l',l''}}=\Gamma_{l',l''}/\Gamma^{max}_{l',l''}$.
The first, second, and third rows correspond to non-interacting $\Delta=0$, strong-interacting $\Delta=-5$ and moderate-interacting $\Delta=-1.5$ systems, respectively.
For all three rows, the evolved time are set as $t=0$, $T_B/4$, $T_B/2$, $3T_B/4$ and $T_B$ from left to right, respectively.
The parameters are chosen as $B=0.05$ and $L_t=101$.
The red regions of $0.25<\overline{\Gamma_{l',l''}}\leq 1$ represent out of range of the colorbar.}
\label{fig:8}
\end{figure*}

In addition to the longitudinal spin-spin correlations, we also calculate the time-dependent two-magnon correlation
\begin{eqnarray}
\Gamma_{l',l''}(t)=\langle\psi(t)|\hat{S}^+_{l'}\hat{S}^+_{l''}\hat{S}^-_{l''}\hat{S}^-_{l'} |\psi(t)\rangle,
\end{eqnarray}
which give the correlations between magnons at sites $l'$ and $l''$. If the two magnons propagate within the region $l\le l_b$, different from the longitudinal spin-spin correlation, there will be complete zero two-magnon correlations for $|l'|>l_b$ or $|l''|> l_b$.
For comparison, we calculate the two-magnon correlations in the BOs of non-interacting magnons, strong-interacting magnons and moderate-interacting magnons, see the first, second and third rows of Fig.~\ref{fig:8}, respectively.
For all the three cases, the evolved time for two-magnon correlations are set as A ($t=0$), B ($t=T_B/4$), C ($t=T_B/2$), D ($t=3T_B/4$) and E ($t=T_B$) from left to right, respectively.
The parameters in the three cases are the same as those in Sec.~\ref{SIIIA}, \ref{SIIIB} and \ref{SIIIC}, respectively.

For non-interacting magnons, the region with non-zero two-magnon correlations expands and shrinks in a Bloch period and is consistent with the region in $|l|<\frac{2}{B}|\sin\frac{Bt}{2}|$ of magnon excitations, see the first row of Fig.~\ref{fig:8}.
For strong-interacting magnons, $\Gamma_{l',l''}(t)$ mainly distributions along two minor off-diagonal lines ($l',l'\pm 1$).
It significantly manifests the formation of magnon BSs.
The two magnons bound together and undergo a fractional BOs with a reduced amplitude $\frac{1}{2\Delta B}$, see the second row of Fig.~\ref{fig:8}.
For moderate-interacting magnons, apart from the clear signal in the off-diagonal $\Gamma_{l',l'\pm 1}(t)$, there also exist fractional distributions in $\Gamma_{l',l''\ne l'\pm 1}(t)$.
It means that the BOs of free-magnon and bounded-magnon components coexist, as shown in the third row of Fig.~\ref{fig:8}.
%

\bibliographystyle{apsrev4-1}


\end{document}